\documentclass[
superscriptaddress,aip,apl,reprint,twocolumn,amsmath,amssymb,showpacs]{revtex4-2}
\usepackage{}

\usepackage[dvipsnames]{xcolor} 

\usepackage{graphicx}  
\usepackage{mathptmx} 
\usepackage{epstopdf}
\usepackage{dcolumn}   
\usepackage{bm}        
\usepackage{amssymb}   
\usepackage{amsmath}   
\usepackage{amsthm}
\usepackage{chemarrow}
\usepackage{color}
\usepackage{mathrsfs}
\usepackage{bbold}
\usepackage{bbm}
\usepackage[a4paper,colorlinks=true,
linkcolor=blue,citecolor=blue,
pdfauthor={ },
pdftitle={ },
pdfsubject={ },
pdfkeywords={ }]{hyperref}

\theoremstyle{plain}
\newtheorem*{theorem*}{Theorem}

\begin{document}

\title{Observation of continuous time crystals and quasi-crystals in spin gases}

\date{\today}

\author{Ying Huang}
\email[]{These authors contributed equally to this work}
\affiliation{
CAS Key Laboratory of Microscale Magnetic Resonance and School of Physical Sciences, University of Science and Technology of China, Hefei 230026, China}
\affiliation{
CAS Center for Excellence in Quantum Information and Quantum Physics, University of Science and Technology of China, Hefei 230026, China}
\affiliation{
\mbox{Hefei National Laboratory, University of Science and Technology of China, Hefei 230088, China}}

\author{Tishuo Wang}
\email[]{These authors contributed equally to this work}
\affiliation{
Guangdong Provincial Key Laboratory of Quantum Metrology and Sensing, and School of Physics and Astronomy, Sun Yat-Sen University (Zhuhai Campus), Zhuhai 519082, China}
\affiliation{
State Key Laboratory of Optoelectronic Materials and Technologies, Sun Yat-Sen University (Guangzhou Campus), Guangzhou 510275, China}

\author{Haochuan Yin}
\affiliation{
CAS Key Laboratory of Microscale Magnetic Resonance and School of Physical Sciences, University of Science and Technology of China, Hefei 230026, China}
\affiliation{
CAS Center for Excellence in Quantum Information and Quantum Physics, University of Science and Technology of China, Hefei 230026, China}
\affiliation{
\mbox{Hefei National Laboratory, University of Science and Technology of China, Hefei 230088, China}}

\author{Min Jiang}
\email[]{dxjm@ustc.edu.cn}
\affiliation{
CAS Key Laboratory of Microscale Magnetic Resonance and School of Physical Sciences, University of Science and Technology of China, Hefei 230026, China}
\affiliation{
CAS Center for Excellence in Quantum Information and Quantum Physics, University of Science and Technology of China, Hefei 230026, China}
\affiliation{
\mbox{Hefei National Laboratory, University of Science and Technology of China, Hefei 230088, China}}

\author{Zhihuang Luo}
\email[]{luozhih5@mail.sysu.edu.cn}
\affiliation{
Guangdong Provincial Key Laboratory of Quantum Metrology and Sensing, and School of Physics and Astronomy, Sun Yat-Sen University (Zhuhai Campus), Zhuhai 519082, China}
\affiliation{
State Key Laboratory of Optoelectronic Materials and Technologies, Sun Yat-Sen University (Guangzhou Campus), Guangzhou 510275, China}

\author{Xinhua Peng}
\email[]{xhpeng@ustc.edu.cn}
\affiliation{
CAS Key Laboratory of Microscale Magnetic Resonance and School of Physical Sciences, University of Science and Technology of China, Hefei 230026, China}
\affiliation{
CAS Center for Excellence in Quantum Information and Quantum Physics, University of Science and Technology of China, Hefei 230026, China}
\affiliation{
\mbox{Hefei National Laboratory, University of Science and Technology of China, Hefei 230088, China}}

\begin{abstract}
Continuous time crystal (CTC) and quasi-crystal (CTQC) are two novel phases of matter characterized by the spontaneous breaking of continuous time-translation symmetry.
To date, realizations of CTCs with periodic oscillations have been reported in only a few physical platforms, and their complex properties still require further exploration.
Additionally, CTQCs, which feature quasi-periodic oscillations at multiple incommensurate frequencies, remain elusive.
Here we report the experimental observation of CTC and CTQC signatures in noble-gas nuclear spins that interact nonlinearly with each other through feedback mechanisms.
The observed limit cycle and quasi-periodic phases display persistent spin oscillations with coherence times extending beyond several hours.
Notably, these oscillations are robust against noise perturbations and exhibit random time phases upon repetitive realization, epitomizing the continuous time-translation symmetry-breaking intrinsic to CTCs and CTQCs.
As the feedback strength increases,
the system undergoes a phase transition into a new phase characterized by chaotic oscillations,
indicative of the ``melting" of time crystals.
Interestingly, within certain feedback regimes,
we even observe an unusual reverse phase transition from a chaotic phase back to time crystal phases.
This work broadens the catalog of new phases of spin gas and unlocks opportunities in precision measurements,
including multimode masers and tests of fundamental symmetries.
\end{abstract}

\maketitle

Time crystals, which exhibit broken time-translation symmetry analogous to the broken space-translation symmetry in ordinary crystals, have captured significant interest from the physics community since their initial proposal.
The original concept of time crystals involved closed systems and demonstrated continuous time-translation symmetry breaking through oscillatory dynamics\cite{wilczek2012quantum,shapere2012classical}.
However, a no-go theorem soon revealed that such time crystals are prohibited by nature\cite{nozieres2013time,bruno2013impossibility,watanabe2015absence},
prompting scientists to explore new possibilities.
One promising direction extends the concept of time crystals into periodically driven systems, known as discrete time crystals\cite{sacha2017time,khemani2019brief,sacha2020time,else2020discrete,zaletel2023colloquium}.
They break the discrete time-translation symmetry imposed by periodic external driving and are characterized by subharmonic oscillations.
Numerous experimental realizations of discrete time crystals have emerged\cite{zhang2017observation,choi2017observation,pal2018temporal,rovny2018observation,kyprianidis2021observation,randall2021many,mi2022time,taheri2022all,wang2022observation,frey2022realization}.
Recently, another approach using open systems to circumvent the no-go theorem has gained attention\cite{iemini2018boundary,buvca2019non,kessler2019emergent}.
This approach allows for continuous time crystals (CTCs) that continuously break time-translation symmetry,
closely aligning with the spirit of the original proposal.
To date, CTCs have been realized on several platforms, including atomic Bose–Einstein condensate\cite{kongkhambut2022observation}, metamaterial nanostructure\cite{liu2023photonic}, semiconductor\cite{greilich2024robust},
polariton condensate\cite{carraro2024solid}, and Rydberg gas\cite{wu2024dissipative}.
Currently, exploring time crystals across a diverse range of physical systems and uncovering their complex properties is at the forefront of scientific research.
Beyond time crystals, another intriguing direction is the exploration of time quasi-crystals\cite{giergiel2019discrete}, which are ordered but not periodic in time, akin to common quasi-crystals.
Recent experiments have demonstrated the existence of discrete time quasi-crystals in quasi-periodically driven systems \cite{he2024experimental} or in periodically driven systems where the induced precession frequency is incommensurate with the drive \cite{autti2018observation}.
However, research on continuous time quasi-crystals (CTQCs) is still pending.

Here we report the experimental observation of continuous time crystals and quasi-crystals in noble-gas nuclear spins.
Our experiments use $^{129}$Xe noble gas as a testbed,
which interacts with overlapping $^{87}$Rb gas.
The $^{87}$Rb gas measures the $^{129}$Xe spin signal and feeds this signal back to $^{129}$Xe through feedback coils.
This feedback mechanism induces nonlinear interactions between $^{129}$Xe atoms,
which is a crucial factor for generating CTCs and CTQCs.
Interestingly,
while magnetic-field inhomogeneities are typically seen as detrimental due to their dephasing effects on spin relaxation,
we intentionally introduce a gradient magnetic field in our setup, which significantly enriches the dynamics of the noble gas.
In this situation, we observe limit cycle, quasi-periodic, and chaotic spin oscillations,
which exhibit signatures of CTCs, CTQCs, and ``melting" of the time crystals, respectively.
The time phases of the limit cycle and quasi-periodic oscillations takes random values for repeated realizations, confirming the  spontaneous breaking of continuous time-translation symmetry\cite{kongkhambut2022observation}.
We investigate phase transitions of the time crystal across system parameters such as feedback strength and magnetic gradient.
The phase diagram shows that the time crystal phase span a wide range of parameters.
Additionally, our time crystal can persist its oscillation pattern against temporal perturbations, verifying the robustness.
Our work paves new paths for realizing and exploring CTCs and CTQCs,
offering exciting opportunities for applications in fields such as quantum metrology.

~\

\noindent
\textbf{\large{Results}}

\begin{figure}[t]  
	\centering
	\makeatletter
	\def\@captype{figure}
	\makeatother
	\includegraphics[width=0.48\textwidth]{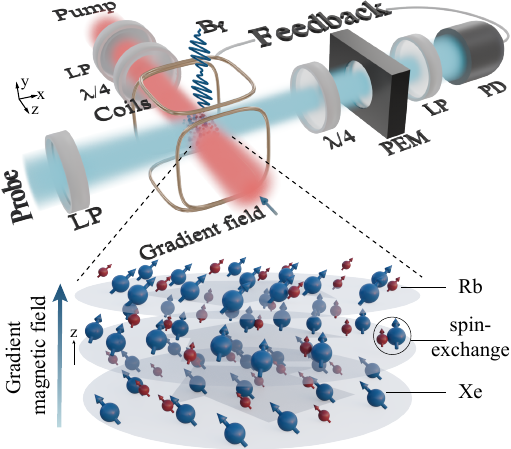}
	\caption{\textbf{Schematic of the experimental setup.} The $^{129}$Xe atoms are polarized and detected through spin-exchange collisions with optically pumped $^{87}$Rb , where $^{87}$Rb atoms are optically polarized by a circularly-polarized pump beam and detected via a rotation of the polarization of a linearly-polarized probe beam. The detected information of $^{129}$Xe spins are magnetically coupled to a feedback circuit, which feeds back real-time $B_y$ through coils along $y$ and induces nonlinear dynamics in $^{129}$Xe spins. The coils along $z$ generates inhomogeneous bias field $B_z$, resulting in continuous Larmor frequency distribution. LP, linear polarizer; $\lambda$/4, quarter-wave plate; PD, photodiode; PEM, photoelastic modulator.}
	\label{fig:setup}
\end{figure}

\noindent
Our system,
exemplified by the Rb-Xe configuration,
integrates overlapping noble gas and alkali-metal gas.
In the setup depicted in Fig.\,\ref{fig:setup},
the $^{129}$Xe nuclear spins are continuously optically pumped through spin-exchange with embedded $^{87}$Rb\,[\onlinecite{walker1997spin},\onlinecite{gentile2017optically}].
These $^{87}$Rb atoms are polarized by a circularly polarized laser beam tuned to the Rb D1 transition at 795\,nm.
A linear gradient magnetic field is applied along the same direction as the Rb pump beam ($z$-axis), resulting a continuous Larmor frequency distribution $\rho(\omega)$.
The $^{129}$Xe nuclear spin component is measured via interactions with the $^{87}$Rb spins using a linearly polarized probe beam oriented along $x$-axis\cite{wu1986optical}.
The intensity change of the probe beam,
which is proportional to the average $^{129}$Xe spin polarization $\bar{P}_x=\int{\rho(\omega)P_x(\omega)\rm{d}\omega}$,
is recorded by a photodiode.
The corresponding $^{129}$Xe signal is fed back to the nuclear spins as a magnetic field along $y$, expressed as $B_{\textrm{f}}=-\alpha\bar{P}_x/\gamma$ where $\alpha$ represents the feedback strength and $\gamma$ is the gyromagnetic ratio of $^{129}$Xe.
More details about our feedback scheme can be found in the Supplementary Information\cite{SI}.

Under this feedback scheme, the system dynamics are governed by a set of nonlinear Bloch equations:
\begin{equation}
\left\{
    \begin{aligned}
    \dot{P}_x(\omega)&=\omega P_y(\omega)+\alpha P_z(\omega)\int\!{\rho(\omega^\prime)P_x(\omega^\prime)\rm{d}\omega^\prime}-\dfrac{P_x(\omega)}{T_2}\\
    \dot{P}_y(\omega)&=-\omega P_x(\omega)-\dfrac{P_y(\omega)}{T_2}\\
    \dot{P}_z(\omega)&=\alpha P_x(\omega)\int\!{\rho(\omega^\prime)P_x(\omega^\prime)\rm{d}\omega^\prime}-\dfrac{P_z(\omega)}{T_1}+R_{\textrm{SE}}
    \end{aligned}
\right.
\end{equation}
where $T_1$($T_2$) is $^{129}$Xe longitudinal (transverse) relaxation time, $R_{\textrm{SE}}$ is the pumping rate through spin exchange collisions with $^{87}$Rb.
The nonlinear term $\alpha P_z(\omega)\int{\rho(\omega^\prime)P_x(\omega^\prime)\rm{d}\omega^\prime}$ and $\alpha P_x(\omega)\int{\rho(\omega^\prime)P_x(\omega^\prime)\rm{d}\omega^\prime}$ introduces global coupling across all frequencies, which could lead to synchronization across different frequencies and the emergence of complex multi-frequency dynamics or chaotic behavior.
When the feedback strength $\alpha$ is weak,
the noble-gas nuclear spins exhibit a normal phase,
with the spins polarized along the bias field due to the relaxation $\{-{P_x(\omega)}/{T_2},-{P_y(\omega)}/{T_2},-{P_z(\omega)}/{T_2}\}$ and pumping $R_{\textrm{SE}}$.
When the nonlinear interaction is sufficiently strong,
three distinct dynamical phases emerge: limit cycle, quasi-periodic and chaos.
The limit cycle phase, corresponds to the synchronization across different frequencies, is characterized by long-lived periodic oscillations, each with a singular, steady frequency.
The quasi-periodic oscillations are superimposed by oscillators of two incommensurate frequencies $\omega_s$ and $\omega_{\tau}$,
where the value of $\omega_{\tau}$ depends on the feedback strength and gradient field \cite{wang2023feedback}. 
When the ratio of $\omega_s/\omega_{\tau}$ is irrational,
these spin oscillations are ordered but not apparently periodic, different from the limit cycle phase.
The application of a gradient magnetic field is crucial for the emergence of this novel spin phase.
To illustrate the different dynamical behaviours of these phases, three representative experimental examples are presented below.
In order to gain deeper insights into the spin dynamics, 
we numerically investigate the stable dynamics of the three corresponding phases.

\begin{figure*}[t]  
	\centering
	\makeatletter
	\def\@captype{figure}
	\makeatother
	\includegraphics[scale=0.99]{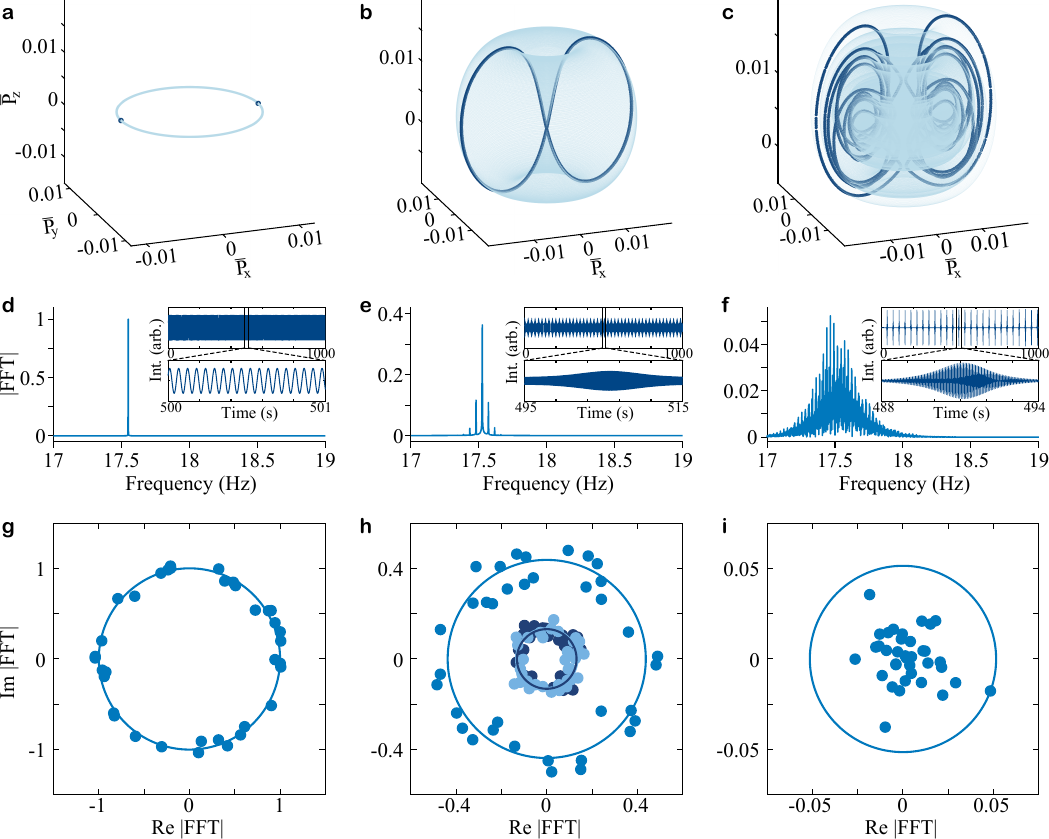}
	\caption{\textbf{Representative dynamical behaviours.} \textbf{a-c}, Phase portraits of the average spin polarization $\{P_x,P_y,P_z\}$ and corresponding Poincaré sections (accentuated) at $P_y=0$ from numerical simulation. \textbf{d-e}, Fourier spectrum (normalized) obtained through the fast Fourier transform (FFT) of Intensity signal recorded by PD in the stationary window (inset). \textbf{g-i}, Relative time phase distribution. Each point represents the Fourier amplitude at the same peak frequency on the complex plane. Each experiment is performed at $\alpha\approx42.52$\,rad/s, $g\approx50$\,nT/cm for the Limit cycle phase; $\alpha\approx159.62$\,rad/s, $g\approx50$\,nT/cm for the Quasi-periodic phase; $\alpha\approx350.86$\,rad/s, $g\approx50$\,nT/cm for the Chaotic phase.}
	\label{fig:2}
\end{figure*}

We numerically simulate the phase portraits of the average spin polarization $\{\bar{P}_x,\bar{P}_y,\bar{P}_z\}$ in three phases.
Additionally, Poincaré sections, defined by the intersections at $P_y=0$, are accentuated in the phase portrait figures to offer a more clear perspective.
The phase portrait of limit cycle is a closed loop and its Poincaré section consists of two discrete points (Fig.\,\ref{fig:2}a).
They both represent a periodic oscillation in which the system revisits the same state after each period.
The phase portrait of quasi-periodic phase traces a torus that never closes, implying the absence of a true period (Fig.\,\ref{fig:2}b).
Correspondingly, the Poincaré section forms closed curves, corroborating the quasi-periodic nature of the system by illustrating the regular yet non-repeating interplay of incommensurate frequencies.
In the chaotic phase, the phase portrait reveals irregular surfaces formed by a dense, scattered distribution of points (Fig.\,\ref{fig:2}c).
The dense distribution of points reflects the underlying chaos,
where minor perturbations in initial conditions induce substantial trajectory divergence.
Additionally, the corresponding Poincaré section exhibits a self-similar pattern of densely scattered points, providing further evidence of the chaotic behavior.
Next, we present the experimentally measured signal of the one-dimensional component $\bar{P}_x$.

\begin{figure*}[t]  
	\centering
	\makeatletter
	\def\@captype{figure}
	\makeatother
	\includegraphics[scale=0.99]{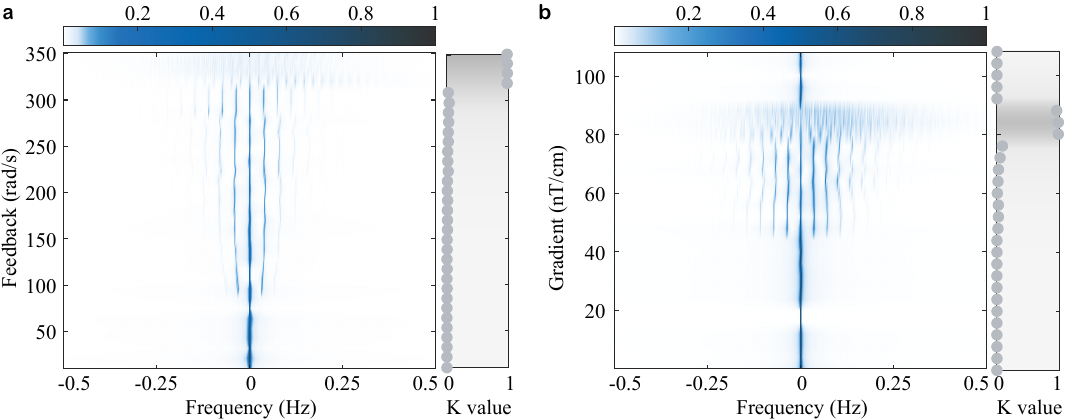}
	\caption{\textbf{Experimental phase diagram and transition.} \textbf{a}, The phase diagram and K dependence on feedback strength $\alpha$ at a fixed magnetic gradient $g\approx50$\,nT/cm. The frequency in phase diagram represents the shift relative to the central frequency. \textbf{b}, The phase diagram and K dependence on magnetic gradient $g$ at a fixed feedback strength $\alpha\approx276.48$\,rad/s.}
	\label{fig:3}
\end{figure*}

Figure \ref{fig:2}d shows a representative time trace of the probe intensity signal in the stationary window,
along with the corresponding spectrum for the limit cycle phase.
The time trace reveals impeccably periodic oscillations, showcasing both a stable amplitude and frequency. 
The basic oscillation pattern of limit cycle is established through characteristic initial transients after the feedback was suddenly switched on, and subsequently maintained in stationary oscillation (see Supplementary Information\cite{SI}).
Note that no pulse or other external stimulus is applied to initiate above dynamic sequence.
By performing a fast Fourier transformation on the data recorded during this interval, we unveil its Fourier spectrum, which is dominated by a single, narrow peak. The linewidth of this peak is constrained solely by the duration of the acquisition time, underscoring the precision and stability of the oscillatory behavior.
These oscillations exhibit extraordinary resilience to temporal perturbations (as discussed below), maintaining their stability even in the presence of a gradient magnetic field applied along the $z$ axis.
Figure\,\ref{fig:2}e illustrates a quintessential example of this quasi-periodic phase.
In this scenario, the Fourier spectrum reveals peaks that are equidistantly spaced.
Overall, the observed nonlinear spin oscillations ordered in time, manifesting as stable limit cycle and quasi-periodic phases, are distinctive signatures of the CTC and CTQC, respectively. Other properties such as the robustness against the perturbations and spontaneous breaking of continuous time-translation symmetry for further confirming the CTC and CTQC are shown in the following.

A typical example of chaotic phase is shown in Fig.\,\ref{fig:2}f.
The temporal signal is composed of wave packets,
but the period and amplitude of these wave packets vary randomly over time.
The corresponding Fourier spectrum features broadened peaks and noisy wings.
To confirm the chaotic nature of this phase,
we employ the Chaos Decision Tree Algorithm\cite{toker2020simple,SI}.
The chaos criteria processes time series recordings and outputs a single value $K$,
which approaches 1 signifies chaotic systems, whereas a value nearing 0 indicates periodic or quasi-periodic systems.
The value of $K$ is approximately 0.9819 for the chaotic phase shown as Fig.\,\ref{fig:2}f.
For comparison, the value of $K$ is approximately 0.0016 for the limit cycle phase as depicted in Fig.\,\ref{fig:2}d, and 0.0028 for the quasi-periodic phase as depicted in Fig.\,\ref{fig:2}e.
The observed chaotic oscillations disrupt the ideal periodicity or quasi-periodicity over time and can be interpreted as the ``melting'' of the time-ordered phases including the CTC and CTQC.

We investigate the spontaneous breaking of continuous time-translation symmetry in CTCs and CTQCs. To confirm that this symmetry breaking occurs spontaneously, the periodic oscillation of noble gas should manifest at any arbitrary initial phase, leading to a random relative time phase across various repeated realizations\cite{kongkhambut2022observation,liu2023photonic}.
To verify that the oscillations in our experimental system exhibit arbitrary time phases, we repeatedly measure the limit cycle and quasi-periodic phases under fixed experimental parameters.
For comparison, the time phase in the chaotic phase is also repeatedly measured.
The measured time phase of the limit cycle is shown in Fig.\,\ref{fig:2}g, displaying a random distribution between $[0,2\pi]$, corresponding to complex Fourier amplitudes around a loop.
This confirms the spontaneous breaking of continuous time-translation symmetry in the limit cycle phase.
The quasi-periodic phase exhibits similar properties, with the complex Fourier amplitude at each frequency randomly distributed on the respective loop, and their time phases covering $[0,2\pi]$ (Fig.\,\ref{fig:2}h).
Our numerical simulations show that the $2\pi$ random phase distribution could arise from initial fluctuations\cite{SI}.

In contrast, the chaotic phase displays distinctly different characteristics, lacking stable oscillation.
In chaotic noble-gas system,
a Fourier peak may not consistently appear in other intervals of a single realization or across different realizations.
As a result, when attempting to calculate the phase of the Fourier amplitude at the same frequency in repeated measurements of chaotic systems, it is often impossible to obtain the same peak's phase for each measurement.
The distribution of the calculated time phase in chaos is shown in Fig.\,\ref{fig:2}i, where all points scatter across a circular surface. This scattering occurs because most points do not correspond to a peak but rather to the broadened baseline or noisy wings.

We now investigate the phase transition as feedback strength increases.
Figure\,\ref{fig:3}a presents the phase diagram derived from the Fourier spectra at various feedback strengths $\alpha$, with a fixed magnetic gradient $g \approx 50$\,nT/cm.
Initially, under weak feedback,
the phase of noble gas remains stable at a no-signal fixed point.
Upon surpassing a critical feedback strength (in our experiment, this threshold is $\alpha\approx9.57$\,rad/s), the limit cycle phase emerges,
although the oscillation amplitude diminishes as  feedback increasing.
As the feedback strength continues to rise, the quasi-periodic phase emerges, revealing an increasing number of harmonics.
For example, at $\alpha\approx265.77$\,rad/s,
around 17 peaks can be distinctly identified, each possessing nearly identical linewidths, significantly below 1 mHz.
The frequency comb gradually stretches with increasing feedback.
With even stronger feedback,
chaotic oscillations manifest, rendering the spectrum blurred.
It is crucial to note that these observations are made in a gradient field;
without this gradient field, neither quasi-periodic nor chaotic phases are observed.

We observe an unusual phase transition from a chaotic phase back to CTC phases.
Previous works only reported the melting of the time crystal into a chaotic phase without noting this reverse transition.
In contrast,
our investigation under various magnetic gradients uncovers this anomalous behavior.
As depicted in Fig.\,\ref{fig:3}b,
the CTC phase undergoes a cycle of formation,
``melting'', and re-formation with increasing magnetic gradient $g$.
Initially, similar to the phase transition pattern observed with varying feedback strength, the periodic oscillation of the noble-gas spins emerges.
As the magnetic gradient $g$ increases, the oscillation amplitude slightly diminishes and the range extends.
Then, chaos abruptly emerges at $g\approx80$\,nT/cm,
indicated by a sudden change in the $K$ value.
However, in a distinct deviation from previous observations, the chaotic phase disappears, followed by the re-emergence of a limit cycle phase above a critical value of $g\approx92$\,nT/cm.
Our finding is significant for the construction of time crystals across a broader range of parameters.

\begin{figure}[t]  
	\centering
	\makeatletter
	\def\@captype{figure}
	\makeatother
	\includegraphics[scale=0.99]{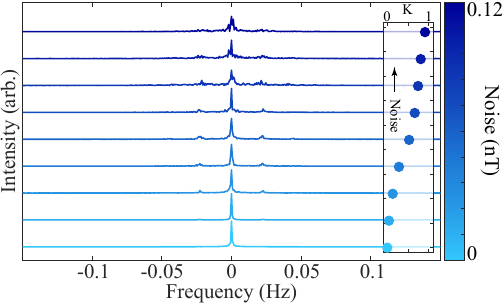}
	\caption{\textbf{Robustness against temporal perturbations.} Fourier spectrum and corresponding $K$ value (inset) of the probe beam intensity signal at increasing noise strength, performed at fixed $\alpha\approx42.52$\,rad/s, $g\approx50$\,nT/cm. The light blue spectrum at the bottom is the noise-free case and the deepening of blue corresponds to the increase in the noise strength. The frequency represents the shift relative to the central frequency.}
	\label{fig:4}
\end{figure}

A unique and attractive feature of time crystal is its robustness.
We demonstrate this robustness by showing the persistence of the oscillation pattern against temporal perturbations.
In experiments, we introduce white noise of varying intensities into the transverse feedback field as a test.
Figure\,\ref{fig:4} illustrates the intensity spectrum at different levels of magnetic noise.
In the low noise regime, the intensity signal retains the fundamental oscillation pattern similar to the noise-free situation.
As the noise strength increases, additional noisy peaks emerge, yet the fundamental oscillation pattern remains clearly recognizable.
With further increases in noise strength, the noise gradually begins to dominate the ordered oscillation.
The corresponding $K$ value (see inset) shows how increasing noise strength ``melts'' the time crystal. 
These observations demonstrate that the observed time crystal can withstand a range of noise levels.
This, combined with the persistence of the time crystal across a wide range of system parameters, confirms the robustness of the observed time crystal.

~\

\noindent
\textbf{\large{Discussions}}

\noindent
Our work broadens the platform for realizing CTCs and CTQCs and enriches the study of their dynamic phases.
We innovatively introduce a gradient magnetic field, significantly diversifying the dynamics of noble gas, including the emergence of quasi-periodic and chaotic phenomena.
The finding of CTQCs would open up an intriguing direction for exploring non-equilibrium phases of matter, both theoretically and experimentally\cite{giergiel2019discrete,autti2018observation,he2024experimental}.
Notably, due to experimental constraints such as limited lifetimes, chaos has rarely been observed in CTC research\cite{kongkhambut2022observation,liu2023photonic,carraro2024solid,wu2024dissipative},
with a recent example documented only in an electron–nuclear spin system in semiconductors\cite{greilich2024robust}.

Beyond its intrinsic conceptual interest,
our work on time crystals holds significant practical value in quantum metrology.
For example, the limit cycle phase and the corresponding precise control of feedback in our CTCs represent a crucial class of spin masers\cite{yoshimi2002nuclear,jiang2021floquet}.
Spin masers, with their long-live persistent oscillations and robustness,
can be widely applied in various fields,
such as high-precision magnetometry\cite{jiang2021floquet}, searches for electric dipole moments\cite{asahi2000fundamental}, tests of fundamental symmetry\cite{heil2013spin}, and probes of new physics\cite{safronova2018search}.
Our spin maser is particularly notable for oscillating at low frequencies (approximately 10\,Hz) while maintaining high relative frequency accuracy.
This makes it ideal for experimental investigations on fundamental questions in physics,
where associated frequencies or frequency shifts are small\cite{heil2013spin}.
Furthermore, existing masers lack diverse dynamic phases, which limits further exploration.
In contrast, the quasi-periodic phase with split frequencies in our CTQCs exhibits richer dynamic behavior, successfully enabling the creation of a multimode maser.
Although demonstrated in the Rb-Xe gas system, we believe that the combination of magnetic gradients with feedback,
as shown in our work,
is suitable for generating multimode masers in many systems, including other alkali and noble-gas pairs, pure alkali-metal gases, and nuclear-spin liquids\cite{suefke2017hydrogen}.

~\

\noindent
\textbf{References}
\bibliographystyle{naturemag}
\bibliography{MSrefs}

~\

\noindent
\textbf{Data availability}.
All study data are included in the article and/or Supplementary Information.

~\

\noindent
\textbf{Acknowledgements}.
This work was supported by the Innovation Program for Quantum Science and Technology (Grant No. 2021ZD0303205), National Natural Science Foundation of China (Grant Nos. T2388102, 11661161018, 11927811, 12004371, 12150014, 12205296, 12274395, and 12261160569), Youth Innovation Promotion Association (Grant No. 2023474), and Chinese Academy of Sciences Magnetic Resonance Technology Alliance Research Instrument and Equipment Development/Functional Development (Grant No. 2022GZL003).
Z.L. was supported by the Guangdong Basic and Applied Basic Research Foundation (Grant No. 2024A1515011406), Fundamental Research Funds for the Central Universities, Sun-Yat-Sen University (Grant No. 23qnpy63), and Guangdong Provincial Key Laboratory (Grant No. 2019B121203005)

~\

\noindent
\textbf{Author contributions}.
Y.H. and T.W. designed and performed experiments, analyzed the data, and wrote the manuscript; H.Y. analyzed the data;
M.J., Z.L. and X.P. proposed the experimental concept, devised the experimental protocols and wrote the manuscript.

~\

\noindent
\textbf{Competing interests}.
The authors declare no competing interests.

\end{document}